\documentclass[final, 3p, 12pt, times]{elsarticle}
\usepackage{color}
\usepackage{amsmath}
\usepackage{amssymb}
\usepackage{graphicx}
\usepackage{hyperref}
\usepackage{slashed}
\usepackage{mathtools}
\usepackage{multirow}

\newcommand{\GeV}{\mbox{GeV}}

\newcommand{\LO}{\mbox{\scriptsize LO}}
\newcommand{\NLO}{\mbox{\scriptsize NLO}}
\newcommand{\mPP}{\mbox{\scriptsize PP}}
\newcommand{\mPV}{\mbox{\scriptsize PV}}
\newcommand{\mVV}{\mbox{\scriptsize VV}}

\definecolor{urlcolor}{RGB}{15,25,112}
\hypersetup{
    bookmarks=true,         
    unicode=false,          
    pdftoolbar=true,        
    pdfmenubar=true,        
    pdffitwindow=false,     
    pdfstartview={FitH},    
    pdftitle={Hadronic production of P-wave charmonia at LHC},    
    pdfauthor= {Likhoded, Luchinsky, Poslavsky},     
    pdfsubject={Next-to-leading order QCD corrections to paired Bc production in e+e- annihilation}, 
    pdfcreator={Likhoded, Luchinsky, Poslavsky},
    pdfproducer={hz}, 
    pdfkeywords={high energy physics, heavy quarkonia, LHC}, %
    pdfnewwindow=true,      
    colorlinks=true,       
    linkcolor=black,          
    citecolor=black,        
    filecolor=magenta,      
    urlcolor=urlcolor           
}

\biboptions{sort&compress}

\journal{Nuclear Physics B}

\begin{document}
\begin{frontmatter}

\title{\Large Next-to-leading order QCD corrections to paired $B_c$ production \\ in $e^+e^-$ annihilation}

\author[sinp]{A.V. Berezhnoy}
\ead{Alexander.Berezhnoy@cern.ch}
\author[ihep,mipt]{A.K. Likhoded}
\ead{Anatolii.Likhoded@ihep.ru}
\author[sinp,jinr,mipt]{A.I. Onishchenko}
\ead{onish@bk.ru}
\author[ihep]{S.V. Poslavsky}
\ead{stvlpos@mail.ru}

\address[sinp]{SINP of Moscow State University, 119991 Moscow, Russia}
\address[ihep]{Institute for High Energy Physics NRC ``Kurchatov Institute'', 142281 Protvino, Moscow Region, Russia}
\address[mipt]{Moscow Institute of Physics and Technology (State University), 141701 Dolgoprudny, Russia}
\address[jinr]{Bogoliubov Laboratory of Theoretical Physics, Joint Institute for Nuclear Research, 141980 Dubna, Russia}

\begin{abstract}
We present theoretical analysis of paired $B_c$ mesons production in $e^+e^-$ annihilation at different energy scales taking into account full next-to-leading order QCD corrections. Both possible electroweak channels are considered: production via virtual photon and via virtual $Z$-boson. We study in detail the role of radiative QCD corrections, which were found to be significant especially at low energies. It is shown that the contribution from $Z$-boson is significant at high energies ($\sqrt{s} > M_Z/2$) especially in the case of paired production of pseudo-scalar and vector ($B_c + B_c^*$) mesons. Azimuthal asymmetry induced by a $P$-violating weak interaction with $Z$-boson is also analyzed. 
\end{abstract}

\end{frontmatter}

\section{Introduction}
In the recent years bound systems of heavy quarks became a perfect laboratory for rigorous theoretical and experimental studies of Standard Model phenomena. Due to the large mass of quarks, production of quark-antiquark pairs can be described by the perturbation theory, while hadronization into observed mesons can be accounted for with the use of effective non-relativistic theories such as NRQCD \cite{Bodwin:1994jh} and pNRQCD \cite{pNRQCD1,pNRQCD2,pNRQCD_Bc}. Among existing heavy quarkonia systems, $B_c$-mesons $(b\bar c)$ are still the most poorly studied experimentally compared to for example charmed or beauty mesons. This is due to the fact that the production cross section is small compared to $(c \bar c)$ or $(b \bar b)$ systems since two pairs of heavy quark pairs must be produced simultaniously. The first observation of $B_c(0^{-})$ was done in hadronic production by CDF collaboration \cite{Abe:1998wi} and then it was extensively studied at Tevatron and LHC. This pseudoscalar $0^{-}$ state is the only one experimentally confirmed $(b\bar c)$ state (recently ATLAS has found evidence for its radial excitation $B_c(2S)$ \cite{Aad:2014laa}). The most significant efforts in the  experimental study of $B_c$ mesons nowadays are undertaking by the LHCb collaboration, which has studied about a dozen of decay channels of $B_c$ meson (see \cite{Egorychev:2015fdl} and references therein) as well as its hadronic production \cite{Aaij:2014ija} which they found in a good agreement with theory \cite{Bcphysics,Bchadronic1,Bchadronic2,Bchadronic3,Bchadronic4}.

Exclusive production of $B_c$ states in $e^+e^-$-annihilation provides a much cleaner setup for the experimental studies of these mesons. There are a number of possibilities to study these mesons considered theoretically such as in $\gamma\gamma$-production \cite{Berezhnoy:1994bb,Berezhnoy:1996an,Berezhnoy:1995ay}, production in Z-decays \cite{Chang:1991bp,Kiselev:1994qp,Yang:2010yg,BcNLO1,BcNLO2} and finally in direct $e^+e^-$-annihilation through virtual $\gamma^*$ exchange \cite{Kiselev:1993iu,Karyasov:2016hfm}. It is worth to mention that study of paired quarkonia production is quite hot research topic, see for example \cite{Berezhnoy:2011xy,pair1,pair2,pair3,pair4a,pair4,pair5,pair6} and references therein. An interest in this subject is partially fueled by the experimental results of collaborations  Belle \cite{Abe:2004ww}, BaBar \cite{Aubert:2005tj}, CMS \cite{Khachatryan:2014iia}, and LHCb \cite{Aaij:2011yc}.

In the present work we study paired production of $S$-wave $(b\bar c)$-mesons through virtual $\gamma^*$ and $Z^*$-boson exchanges  in $e^+e^-$-annihilation taking into account full next-to-leading order (NLO) QCD radiative corrections. We show that these one-loop corrections give sensitive contribution to the whole production cross sections. Additionally, theoretical errors induced by the choice of the renormalization scale in strong coupling $\alpha_S(\mu)$ become several times smaller at NLO compared to leading order (LO). In the case of the production of a pair of pseudoscalar mesons there is an interesting feature related to the cross section becoming zero (both within LO and NLO) at some value of center of mass energy near threshold, which can be potentially interesting for experimental search and study of $B_c^*$ mesons, since all $(b \bar c)$-mesons produced exclusively at this point are $1^-$. We show that the account for virtual $Z^*$-boson exchange changes significantly the behavior of $B_c + B_c^*$ production cross section, in particular it changes the asymptotic behavior at large energies and completely changes the cross section dependence on the renormalization scale $\mu$. Such effects are absent in the case of $B_c + B_c$ and $B_c^* + B_c^*$ production. Parity violation induced by weak interaction is shown to be significant at the energies near $m_Z/2$ for the production cross section of $B_c + B_c^*$ and $B_c^* + B_c^*$, while in the case of $B_c + B_c$ there is no parity violation in all energy range since final state with such quantum numbers can not be produced from the axial part of $Z$-current and thus there is no V-A interference.

\section{The method}

To calculate paired $B_c^{(*)}$ production cross sections we employed a systematic framework offered by NRQCD factorization \cite{Bodwin:1994jh}. This way cross sections are being expressed as the sum over $b\bar c$ and $c\bar b$ channels of products of hard perturbative $b\bar c$, $c\bar b$ cross sections and soft nonperturbative NRQCD matrix elements.

\subsection{NRQCD factorization and projectors}

Production and bound state dynamics of quarkonia involves several different energy scales. First, we have hard scales given by heavy quark masses $m_b$ and $m_c$. Next, there are so called soft and ultrasoft scales. The soft scale is given by the relative momentum of heavy quark-antiquark pair in the quarkonium rest frame $|{\bf p}|\sim m v~(\sim 1/r)$. Here $v$ is the relative velocity of quark-antiquark motion, $r$ is the radius of quarkonium and $m$ is the reduced mass, which in the case of $B_c$ meson is given by $m = \frac{m_c m_b}{m_c + m_b}$. The ultrasoft scale is given by kinetic energy of quark-antiquark motion inside quarkonium $E\sim m v^2$ and of course we have nonperturbative confinement scale $\Lambda_{QCD}$. The nonrelativistic QCD (NRQCD) is the effective theory obtained by integrating out hard degrees of freedom from QCQ. The general hierarchy of scales for quarkonia production and bound state dynamics is $m_b,m_c \gg  m v, m v^2, \Lambda_{QCD}$, as $v \ll 1$.  First, it sets a firm ground for constructing a self consistent effective theory for bound state dynamics and, second, it allows us to write down production cross sections for quarkonia in a factorized form:  hard perturbative production of heavy quark pairs and subsequent soft nonperturbative formation of quarkonia bound states out of these heavy quark-antiquark pairs.              

To calculate matrix elements for $e^+e^-\to B_c^{+ (*)} (P_1) +  B_c^{- (*)} (P_2)$ production processes, we start from the matrix element for $e^+e^-\to c (p_c)\bar b (p_{\bar b}) + b (p_b)\bar c (p_{\bar c})$ with heavy quarks and antiquarks on their mass shells: $p_{Q}^2 = p_{\overline Q}^2 = m_Q^2$. For each of the $B_c$ mesons the momenta of heavy quarks are written as
\begin{eqnarray}
&p_c = \frac{m_c}{m_b+m_c} P_1 + q_1, \quad &p_{\bar b} = \frac{m_b}{m_b+m_c} P_2 - q_1 , \nonumber \\
&p_b = \frac{m_b}{m_b+m_c} P_2 + q_2, \quad &p_{\bar c} = \frac{m_c}{m_b+m_c} P_2 - q_2 ,
\end{eqnarray}
where $q_1, q_2$ are relative momenta of heavy quark-antiquark motion inside $B_c^{+ (*)}$ and $B_c^{- (*)}$ mesons correspondingly. The relative momenta are subject to constraints $q_1\cdot P_1 = 0$ and $q_2\cdot P_2 = 0$. The leading contribution in Fock state expansions of $B_c$-mesons is given by color singlet $S$-wave components. The corresponding  covariant projectors for color-singlet spin-singlet and spin-triplet states are given by 
\begin{eqnarray}
& v (p_{\bar b}) {\bar u} (p_c) = \frac{1}{2\sqrt{2}} \left(\slashed{P}_1 - m_c - m_b\right) \left\{\gamma_5,\, \slashed{\epsilon}_1 \right\}\otimes \left(\frac{1}{\sqrt{N_c}}{\bf 1}\right), \nonumber \\
& v (p_{\bar c}) {\bar u} (p_b) = \frac{1}{2\sqrt{2}} \left(\slashed{P}_2 - m_c - m_b\right)\left\{\gamma_5,\, \slashed{\epsilon}_2 \right\}\otimes \left(\frac{1}{\sqrt{N_c}}{\bf 1}\right) ,
\end{eqnarray}
where $\epsilon_1$, $\epsilon_2$ are spin polarization vectors for $B_c^{+ *}$ and $B_c^{- *}$ mesons, satisfying constraints: $\epsilon_{1}\cdot\epsilon_{1}^{*} = \epsilon_{2}\cdot\epsilon_{2}^{*} = -1$ and $P_1\cdot\epsilon_1 = P_2\cdot\epsilon_2 = 0$.  

Next, within NRQCD factorization as we already mentioned the matrix elements for the production of $B_c^{(*)}$ pairs could be written as 
\begin{eqnarray}
&&\mathcal{M} \left[B_c + B_c\right] = \frac{1}{N_c (m_c + m_b)} \langle O\rangle_{B_c} \mathcal{A}^{PP}, \nonumber \\
&&\mathcal{M} \left[B_c^{*} + B_c\right] = \frac{1}{N_c (m_c + m_b)} \langle O\rangle_{B_c}^{1/2} \langle O\rangle_{B_c^{*}}^{1/2} \mathcal{A}^{VP}_{\mu}\epsilon^{\mu},  \nonumber \\
&&\mathcal{M} \left[B_c^{*} + B_c\right] = \frac{1}{N_c (m_c + m_b)} \langle O\rangle_{B_c^{*}} \mathcal{A}^{VV}_{\mu\nu}\epsilon_1^{\mu}\epsilon_2^{\nu}, 
\end{eqnarray}
where $\mathcal{A}^{PP}$, $\mathcal{A}^{VP}$, $\mathcal{A}^{VV}$ are hard production cross sections of two heavy quark-antiquark pairs projected on the heavy quark-antiquark states with zero relative momenta and $B_c^{(*)}$ quantum numbers with the help of projectors above.  The NRQCD matrix elements $\langle O\rangle_{B_c}$ and $\langle O\rangle_{B_c^{*}}$ are vacuum-saturated analogs of the NRQCD matrix elements $\langle O (^1 S_0)\rangle$ and $\langle O (^3 S_1)\rangle$ for annihilation decays defined in \cite{Bodwin:1994jh}. At present time, these nonperturbative matrix elements can not be derived in the explicit form theoretically (and there is no enough experimental data to extract them from the experiment unlike e.g. for charmonium  \cite{Bolzoni:2013tca,Butenschoen:2012qr,Butenschoen:2011yh,Butenschoen:2009zy,Butenschoen:2010rq,Likhoded:2014kfa,Likhoded:2012hw}), so in this work we will assume that $\langle O\rangle_{B_c^{*}} = \langle O\rangle_{B_c}$ and use the expression of the latter in terms of $B_c$ leptonic constant: $\langle O\rangle_{B_c} = \frac{m_c+m_b}{2} f_{B_c}^2$, which is known at a moment with two loop accuracy \cite{Bcleptonicconstant,KiselevLeptonicConstant,Bcleptonicconstant2}, see also \cite{NLOBcsumrules} for NLO determination of $B_c$ leptonic constant from QCD sum rules.

\subsection{LO and NLO cross section calculation}

\begin{figure}
	\centerline{\includegraphics[width=\textwidth]{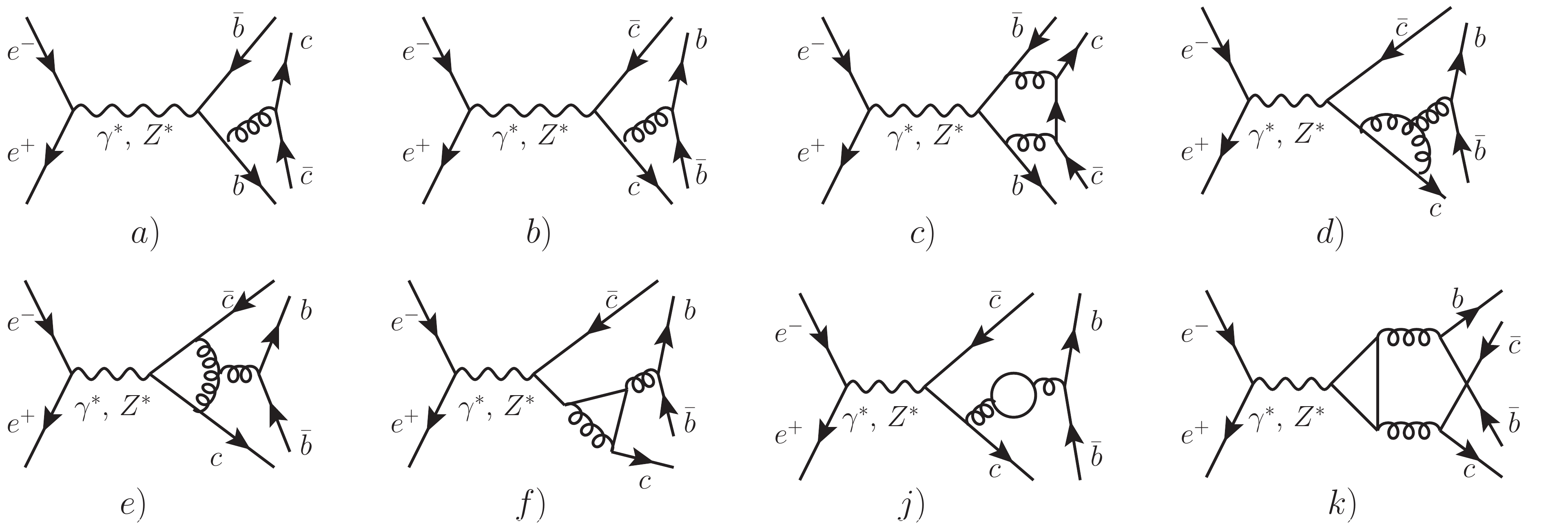}}
	\caption{Typical Feynman diagrams for the process $e^+e^- \to B_c^+ + B_c^-$. Diagrams $a)-b)$ for the tree level, diagrams $c)-k)$ --- sample one-loop diagrams. \label{fig_diags}}
\end{figure}

So, to calculate production cross sections of $B_c^{(*)}$ pairs we need to know expressions for hard production cross sections of two heavy quark-antiquark pairs projected on states with given  quantum numbers. This problem requires automation already at tree level, where we have 8 diagrams, see Fig.~\ref{fig_diags}. To generate Feynman diagrams and obtain analytical expressions for them we used \texttt{FeynArts} package \cite{Hahn:2000kx}. The subsequent processing, such as color and Dirac algebra were performed using several different strategies based on usage of different computer algebra systems such as \texttt{Mathematica} with \texttt{FeynCalc} \cite{FeynCalc1,FeynCalc2} package , \texttt{FORM} \cite{Vermaseren:2000nd} or \texttt{Redberry} \cite{Bolotin:2013qgr}. This way, for example, for $\mathcal{A}_{\mu}^{VP}\epsilon_1^{\mu}$ matrix element accounting only for virtual photon exchange we get  
\begin{eqnarray}
\mathcal{A}_{\mu}^{VP}\epsilon_1^{\mu} = \frac{2 C_F e^2 g_s^2 (m_c + m_b) (3 r^3-3 r^2 + 3 r - 1)}{3 (r-1)^3 r^3 s^3}
\bar u (-k_2)\gamma^{\mu} u (k_1) \epsilon_{\mu\epsilon_1 P_1 P_2},
\end{eqnarray}
where $k_1, k_2$ are incoming momenta of electron-positron pair, $s = (k_1+k_2)^2$, $r = \frac{m_c}{m_c+m_b}$ and $C_F = \frac{N_c^2-1}{2 N_c}$. To get an expression for cross section this matrix element should be squared, summed over $B_c^{*}$ meson polarizations, averaged over polarizations of initial electron-positron pair and integrated over final phase space. In \ref{appendixA} we have gathered the results for leading order differential $B_c^{(*)} + B_c^{(*)}$ cross sections including contributions both from  virtual $\gamma^*$ and  $Z^*$-boson exchanges. It should be noted, that the results for  the case of $\gamma^*$ contribution have already appeared in literature \cite{Kiselev:1993iu}, while account for $Z^*$-boson exchange is a new result.   

The NLO calculation is more involved. At this order we have 172 nonzero diagrams, see Fig.~\ref{fig_diags} for some sample diagrams. To regularize both UV and IR singularities we used dimensional regularization ($D = 4-2\epsilon$). In order to calculate hard perturbative production cross section we used so called method of regions \cite{MethodRegions}. This way we may directly deduce NRQCD short-distance coefficients (hard production cross sections in our case) by setting relative momenta of heavy quark-antiquark motion inside quarkonia to zero before carrying out loop integrations rather than resorting to much more expensive matching calculations. The diagram calculation itself was again performed using several computational strategies. To reduce diagrams to master integrals we used Laporta algorithm \cite{Laporta} and its implementation in \texttt{FIRE} \cite{FIRE1,FIRE2}. Before this step however one should resolve a possible linear dependence of propagators and perform partial fractioning when needed. For that purpose we used \texttt{\$Apart} \cite{Feng:2012iq} package. The tensor integrals were treated using Passarino-Veltman reduction \cite{Passarino-Veltman}. Another strategy employed our own private \texttt{Mathematica} code, which performed topological analysis of the generated diagrams and produced \texttt{FORM} rules for diagram reduction to canonical prototypes involving numerator simplification, partial fractioning and Passarino-Veltman reduction together with subsequent reduction of canonical prototypes to master integrals using reduction database generated with \texttt{FIRE}.  For numerical evaluation of master integrals we used \texttt{PackageX} \cite{Patel:2015tea}. In the case of two-point $A_0$ and $B_0$ master integrals it was required to hold terms up to $\epsilon^1$ order, so we used known analytical results which are given for reader's convenience in \ref{appendixB}. For renormalization procedure we employed so called on-shell scheme\footnote{The gauge coupling renormalization was performed in the modified minimal subtraction $\overline{MS}$ scheme.}:
\begin{eqnarray}
Z_m^{OS} &=& 1 - \frac{3 g_s^2}{16\pi^2} C_F C_{\epsilon} \left[
\frac{1}{\epsilon_{UV}} + \frac{4}{3}\right] + \mathcal{O}(\alpha_s^2), \nonumber \\
Z_2^{OS} &=& 1 - \frac{g_s^2}{16\pi^2} C_F C_{\epsilon}
\left[\frac{1}{\epsilon_{UV}} + \frac{2}{\epsilon_{IR}}
+ 4 \right] + \mathcal{O}(\alpha_s^2), \nonumber \\
Z_g^{\overline{MS}} &=& 1 + \frac{g_s^2}{16\pi^2} \left(
-\frac{11}{6}C_A + \frac{1}{3}n_f\right)
\left[
\frac{1}{\epsilon_{UV}} - \gamma_E + \ln (4\pi)
\right]  + \mathcal{O}(\alpha_s^2) ,
\end{eqnarray}
where $C_F = \frac{N_c^2-1}{2 N_c}$, $C_A = N_c$, $C_{\epsilon} = \left(\frac{4\pi\mu^2}{m^2}e^{-\gamma_E}\right)^{\epsilon}$, $m$ is the heavy quark pole mass, $n_f$ - number of fermions taken into account in gluon self-energies and $\gamma_E$ - Euler's gamma constant. The resulting expressions for renormalized one-loop cross sections in terms of master integrals could be found in supplementary material. For $\gamma_5$ treatment we used two different prescriptions, namely West \cite{West} and Larin \cite{Larin} schemes . It turns out, that for the production of a pair of pseudoscalar (PP) mesons and production of vector and pseudoscalar meson (VP) both schemes give identical results. At the same time in the case of the production of a pair of vector mesons (VV) both schemes gave slightly (less then a percent) different results. In this case we regard this discrepancy as an error related to $\gamma_5$ scheme dependence. See \cite{gamma5} for discussion of problems present when dealing with $\gamma_5$ in dimensional regularization. It is also worth to mention that if we consider only $\gamma^*$ contribution both schemes give identical results for all final states.

\section{Results}

\noindent
To present numerical estimates we used the following values of
parameters:
\begin{gather*}
f_{B_c}^2 = 0.16\, \GeV^3, \quad m_c = 1.5\,\GeV, \quad m_b = 4.8\,\GeV, \quad m_t = 174.6\,\GeV,\\
m_Z = 91.2\,\GeV, \quad \Gamma(Z) = 2.5\,\GeV,  \quad \sin^2(\theta_W) = 0.23.
\end{gather*}
The running of strong coupling constant was taken at one-loop accuracy and is given by
$$
\alpha_S(Q) = \frac{2\pi}{\left(11 -2/3 n_f\right) \ln\left(Q/\Lambda \right)},
$$
with $\alpha_S(m_Z) = 0.1184$. The renormalization scale $\mu_R$ and strong coupling scale $Q$ were chosen as\footnote{See the discussion below.} $\mu_R = Q = \mu = \sqrt{s}$.

\begin{figure}
\centerline{\includegraphics[width=0.45\textwidth]{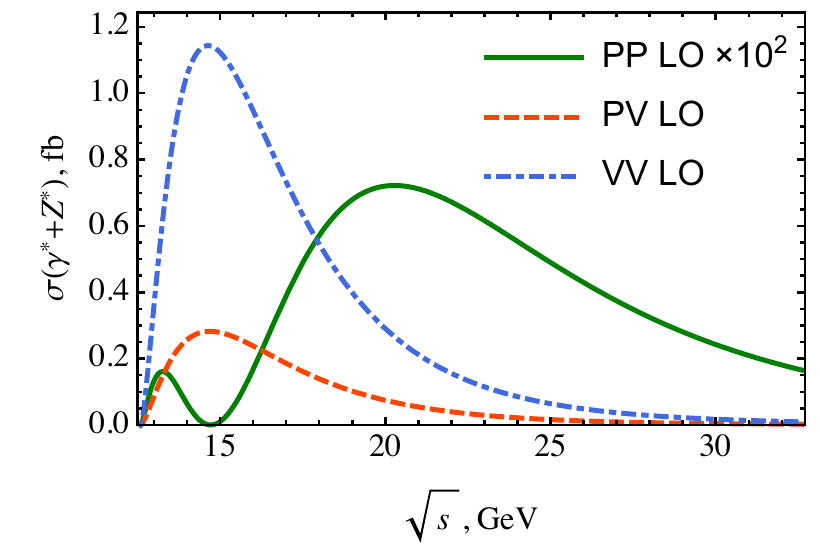} \hspace{0.02\textwidth} \includegraphics[width=0.45\textwidth]{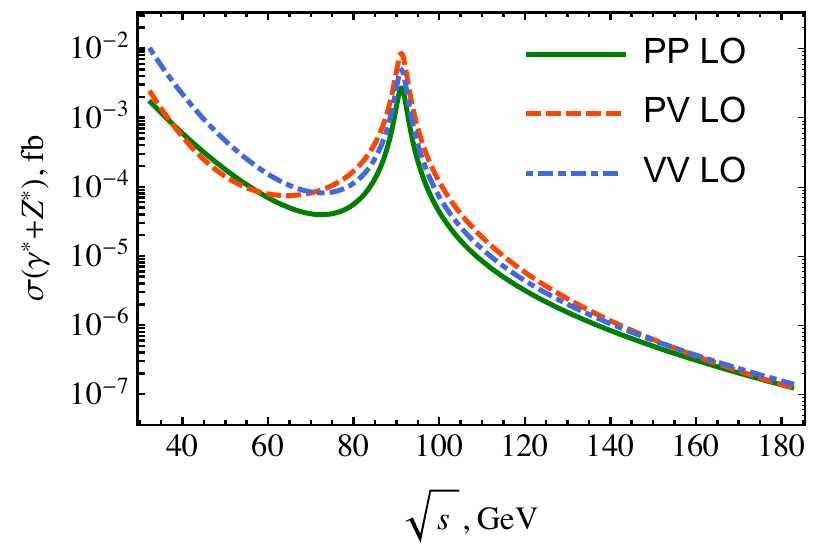}}
\caption{Tree level cross section dependence on $\sqrt{s}$ at low (left) and high (right) energies: solid curve for a pair of pseudoscalar states (PP) raised by a factor of 100, dashed for pseudoscalar plus vector (PV) and dot-dashed for a pair of vector mesons (VV).
\label{fig_tree_ecm}}
\end{figure}

Fig.~\ref{fig_tree_ecm} shows tree level cross sections near production threshold (left) and near $Z$-boson peak (right). First of all, from the left plot we see that the production cross section of a pair of pseudoscalar (PP) mesons $e^+ + e^-\to B_c^+ + B_c^-$ is zero at some center of mass energy near threshold. This point is 
\begin{equation}
\label{eq_zero_PP_point}
s_{\mPP}  = 2 M^2  \frac{\left(3 r^4-4 r^3+6 r^2-4 r+1\right)}{r (1-r) \left(3 r^2-2 r+1\right)}, \qquad \sqrt{s_{\mPP}} \approx 14.7\,\GeV,
\end{equation}
where $r = m_c/M$ and $M = m_c + m_b$. If we neglect $Z$-boson contribution then the cross section at this point is exactly zero. Accounting for $Z$-boson it is given by:
\begin{multline*}
\sigma_{\mbox{\scriptsize PP}}(s_{\mbox{\scriptsize PP}}) = 
\pi^3 \alpha^2 \alpha_S^2 f_{B_c}^4 M^4 \left(1-\frac{4 M^2}{s_{\mPP}}\right)^{3/2} \times 
\frac{2 (2 r-1)^2}{243 (r-1)^2 r^2 \left(3 r^2-2 r+1\right)^2} \times \\ \times 
\frac{   \left(-2 \cos \left(2 \theta_W\right)+\cos \left(4 \theta_W\right)+2\right) \csc^4\left(\theta_W\right) \sec^4\left(\theta_W\right)}{ s_{\mPP}^3 \left(\left(m_Z^2-s_{\mPP}\right){}^2+\Gamma^2 m_Z^2\right)}
\approx 
2.7 \times 10^{-6} \,\, \mbox{fb},
\end{multline*}
which is negligibly small (as it is generally the case for $Z$-boson contribution at small energies). 

At the same time, production cross sections of $B_c^+ + B_c^{*-}$ (PV) and $B_c^{*+} + B_c^{*-}$ (VV) mesons have maxima near $\sqrt{s} \approx \sqrt{s_{\mPP}} \approx 15 \,\GeV$ (moreover $\sigma_{\mVV}/\sigma_{\mPV} \approx 4$ at this energy). So, one can use this energy point as a some kind of ``$B_c^*$-factory'' since all $(b \bar c)$-mesons produced exclusively at this point are $1^-$. While there is no yet any experimental evidence for $B_c^*$, such setup can be potentially interesting for experimental search and study of $B_c^*$ mesons.

At high energies the cross sections for all considered processes tend to zero as $1 / s^3$. However, without the account of $Z$-boson contribution the production cross section of pseudo-scalar plus vector (PV) mesons goes to zero as $1 / s^4$. This can be explained by the fact that PV production is forbidden in massless limit due to chirality conservation (each quark-antiquark pair is produced from either photon or gluon which both are $1^-$ and there is no way to combine them into $1^- \,+\, 0^-$  state combination). In the case of massive quarks it is additionally suppressed by $M/\sqrt{s}$ factor giving extra power of $s$ to the asymptotic behavior of cross section. Accounting for $Z$-boson contribution containing axial current this restriction is removed and $\sigma_{\mPV} \to 1/s^3$ at high energies. In the case of a pair of pseudoscalar (PP) and a pair of vector (VV) mesons an account of $Z$-boson contribution raises cross section at high energies by a factor:

\begin{multline*}
\frac{\sigma_{\mbox{\scriptsize PP, VV}}(Z^* + \gamma^*)}{\sigma_{\mbox{\scriptsize PP, VV}}(\gamma^*)}
 \underset{s \gg M_{Bc}}{\longrightarrow} 
 \frac{\csc ^4\left(2 \theta _W\right)}{8 (r (3 r-2)+1)^2} \left(\left(2 r \left(r \left(9 r^2-6 r+5\right)-2\right)+1\right) \cos \left(4 \theta _W\right)
 +\right. \\ \left. +
 \left(2-4 r \left(18 r^3-6 r^2+r+2\right)\right) \cos \left(2 \theta _W\right)+6 (r (3 r-1)+1) (r (5 r-3)+1)\right) \approx 1.5.
\end{multline*}

\begin{figure}
\begin{center}
\includegraphics[width=0.45\textwidth]{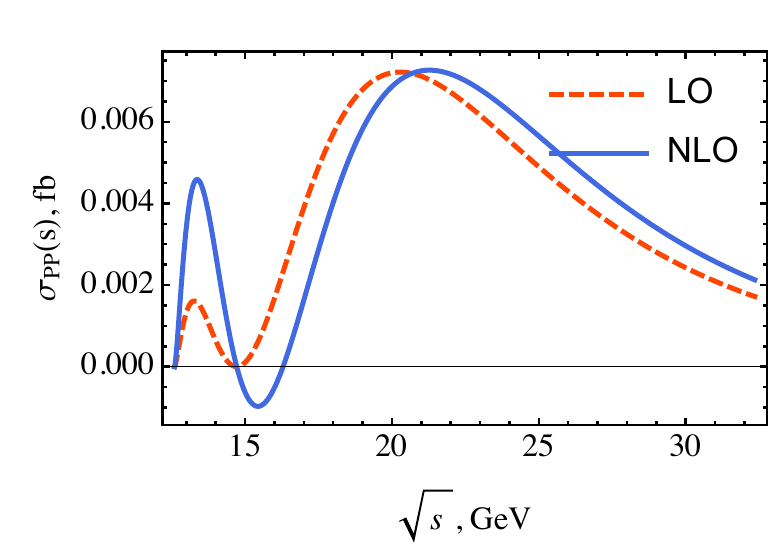} 
\hspace{0.02\textwidth} 
\includegraphics[width=0.45\textwidth]{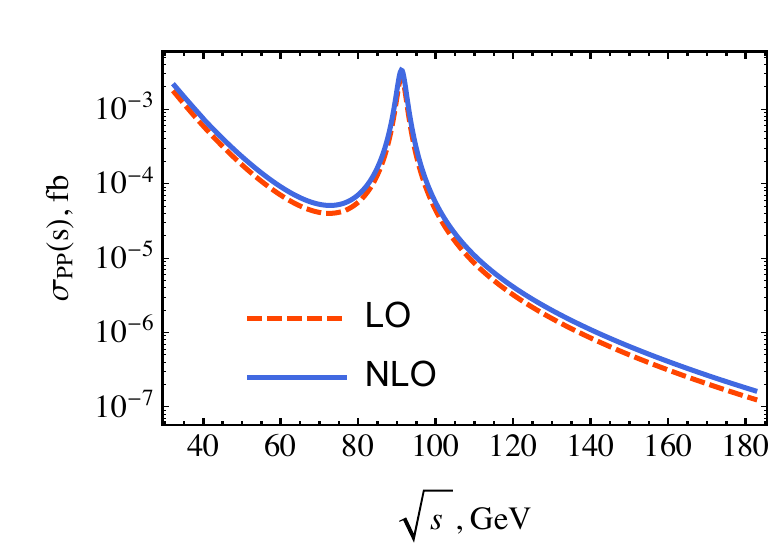}  \\
\includegraphics[width=0.45\textwidth]{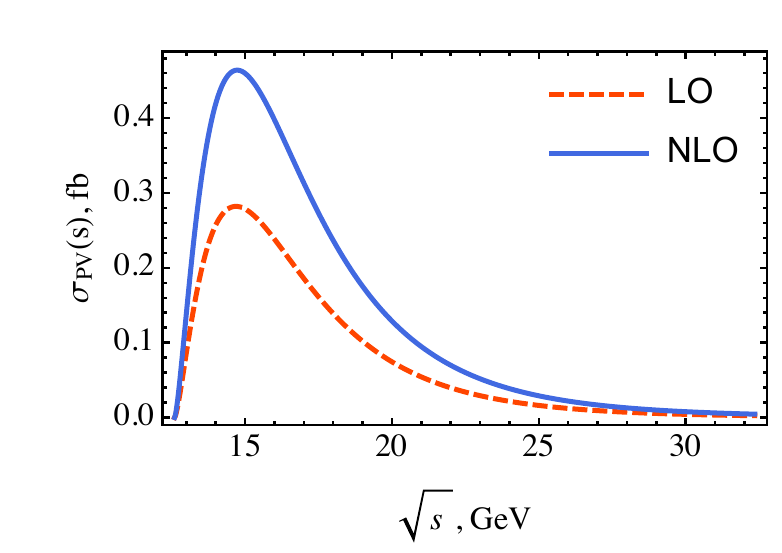} 
\hspace{0.02\textwidth} 
\includegraphics[width=0.45\textwidth]{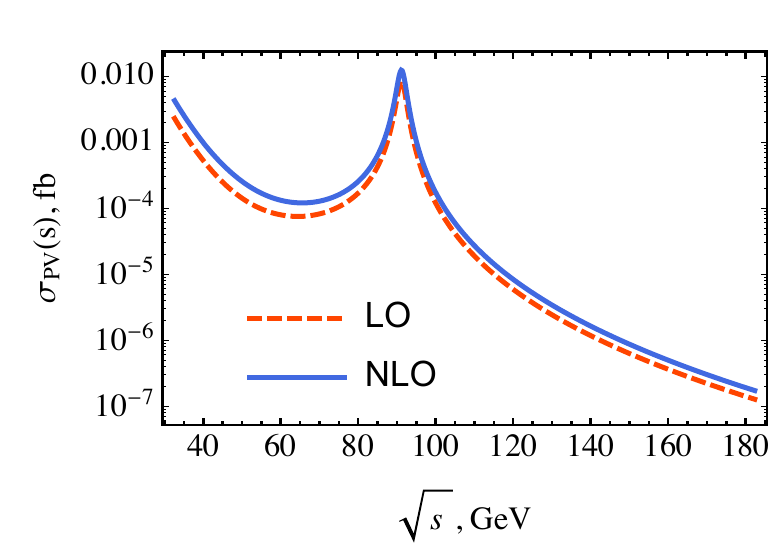}  \\
\includegraphics[width=0.45\textwidth]{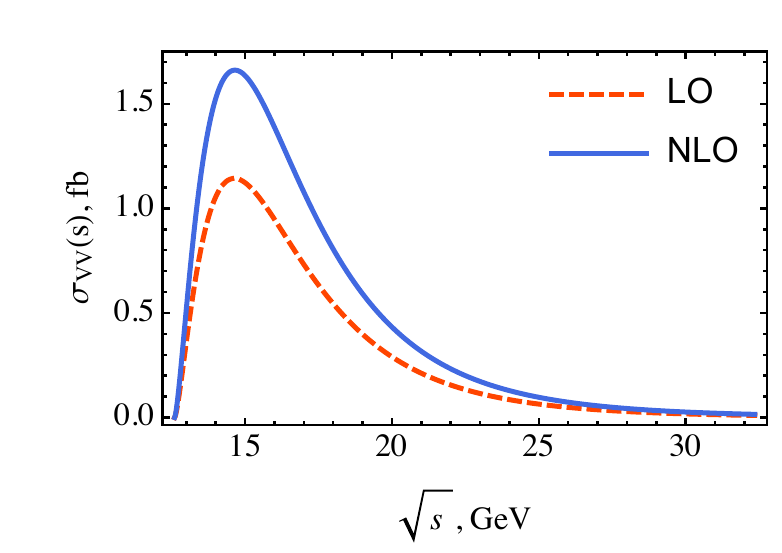} 
\hspace{0.02\textwidth} 
\includegraphics[width=0.45\textwidth]{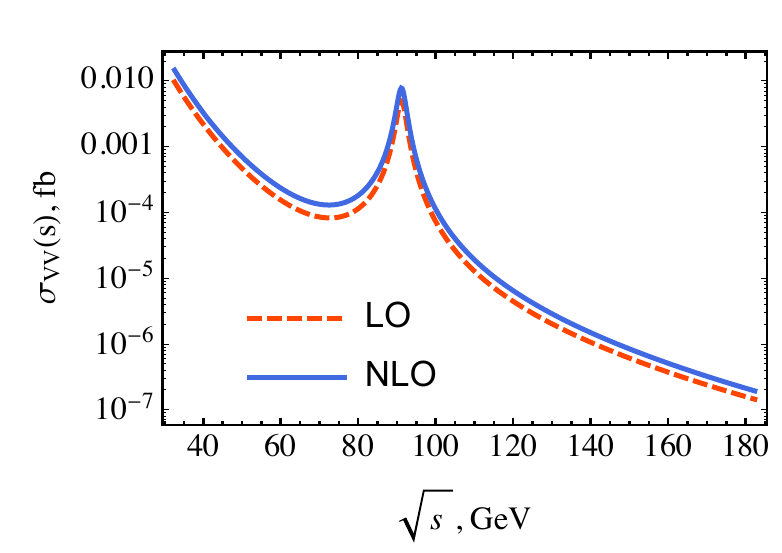} 
\hspace{0.02\textwidth}
\end{center}
\caption{NLO cross sections dependence on $\sqrt{s}$ at low energies near production threshold (left) and high energies near $Z$-boson peak (right). Solid curve for a full NLO cross section, dashed for LO contribution. Upper plots correspond to production of a pair of pseudoscalar (PP) mesons, middle for pseudoscalar and vector (PV) and lower for a pair of vectors (VV).
\label{fig_nlo_ecm}}
\end{figure}

\begin{figure}
\centerline{\includegraphics[width=0.45\textwidth]{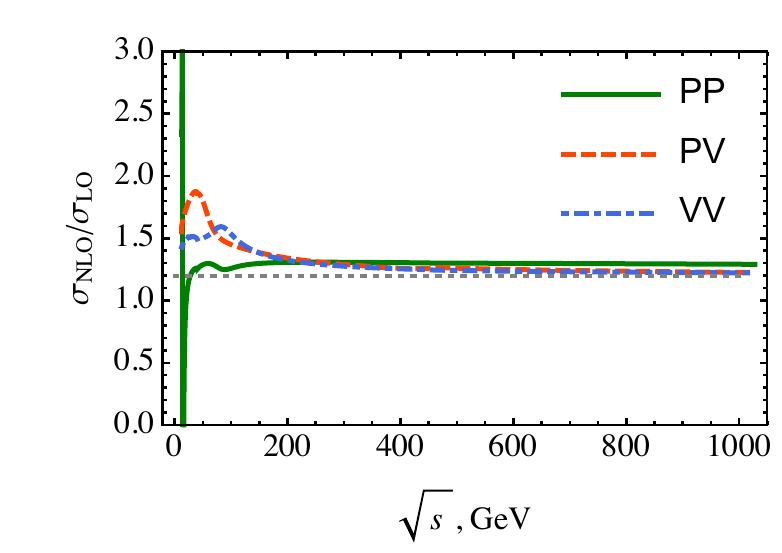}}
\caption{Ratio of full NLO and LO cross sections dependence on $\sqrt{s}$. Solid curve for a pair of pseudoscalar states (PP), dashed for pseudoscalar plus vector (PV), and dot-dashed for a pair of vector mesons (VV). Horizontal dotted line corresponds to $\sigma_{\NLO}/\sigma_{\LO} = 1.2$. Sharp PP behaviour at low energies is a consequence of LO cross section going to zero at $s_{\mPP}$ as defined in \eqref{eq_zero_PP_point}.
\label{fig_kfactor}}
\end{figure}

Now let us turn to one-loop corrections. Fig.~\ref{fig_nlo_ecm} shows NLO cross sections at different energy scales, while  Fig.~\ref{fig_kfactor} shows the ratio $\sigma_{\NLO}/\sigma_{\LO}$. 
It is seen, that in general loop corrections are more important at low energies and may even raise the cross sections by a factor of two as it is the case for PV mesons production. At high energies $\sigma_{\NLO}/\sigma_{\LO}$ tends to $\approx 1.2$. Next, from Fig.~\ref{fig_nlo_ecm} it is seen that in the PP case loop contribution is negative at low energies (square of one-loop contribution is discarded at $O(\alpha_S^3)$). This leads to unphysical result near $s_{\mPP}$ point defined before in \eqref{eq_zero_PP_point}, where tree cross section goes to zero. So, to correctly describe production cross section at this point at least two-loop calculation is required as well as account for relativistic  $\mathcal{O} (v^2)$ corrections (such corrections to paired $B_c$-production were recently studied in \cite{Karyasov:2016hfm}).

\begin{figure}
\begin{center}
\includegraphics[width=0.45\textwidth]{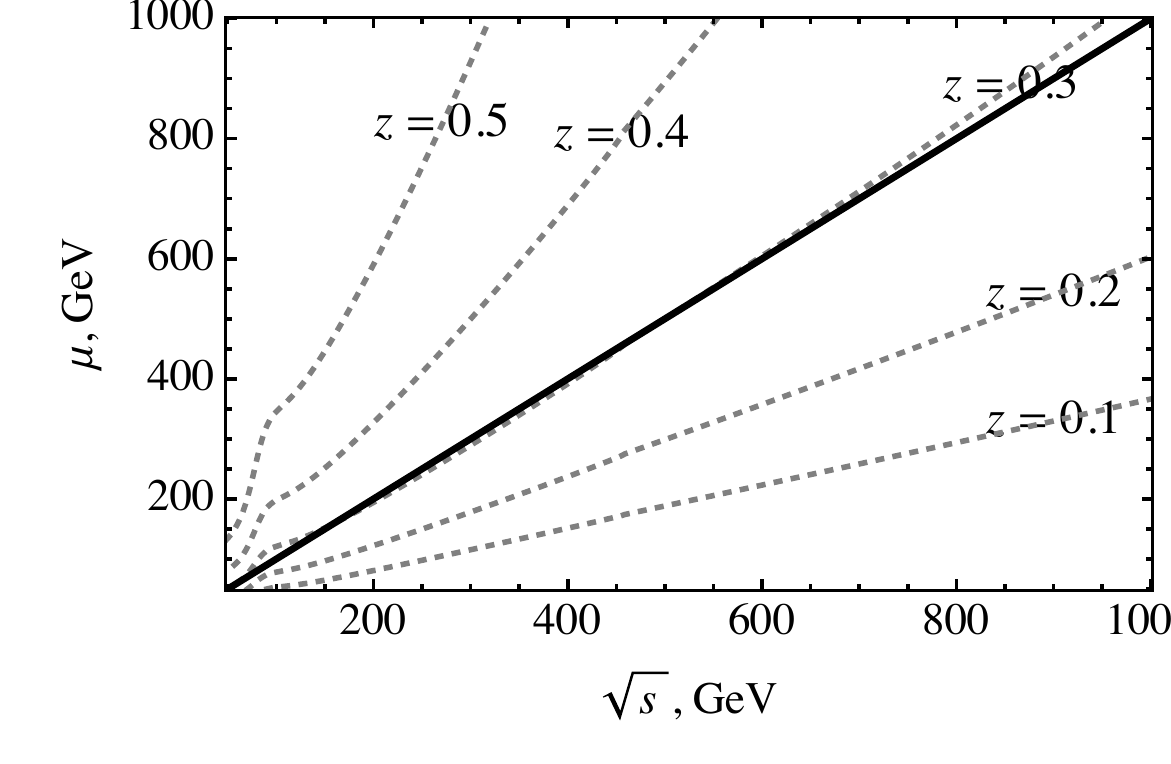} 
\hspace{0.02\textwidth} 
\includegraphics[width=0.45\textwidth]{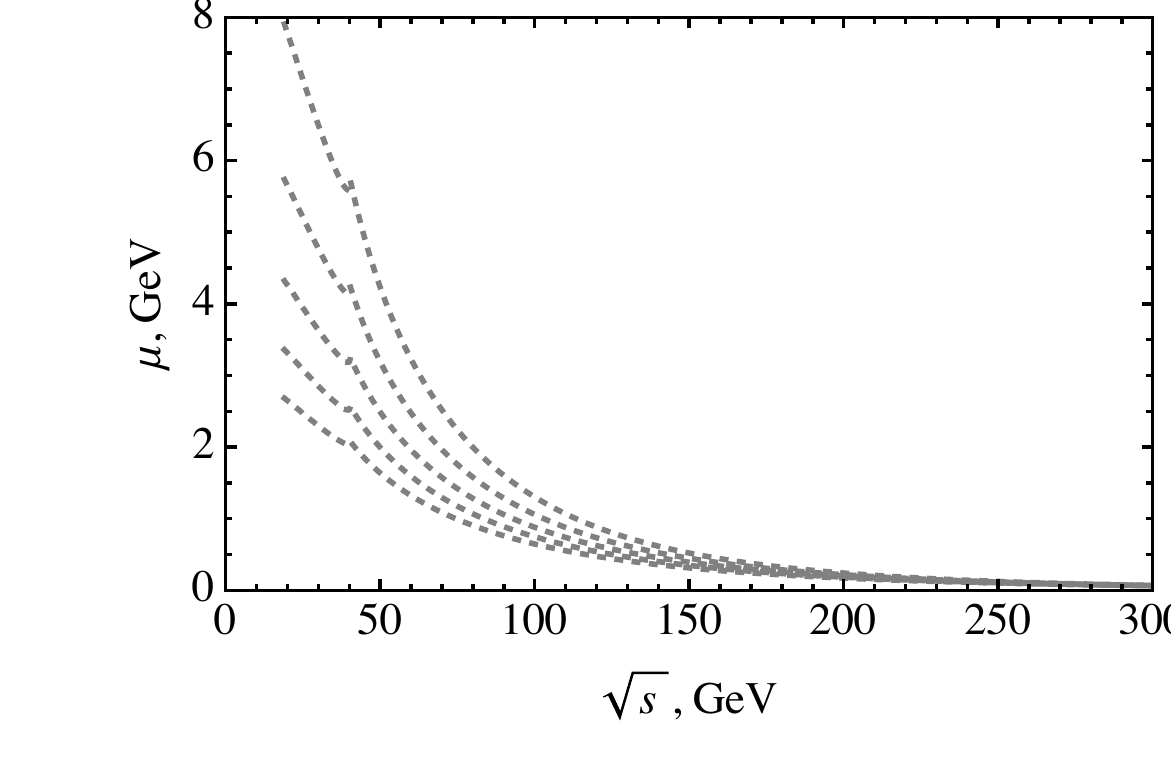}  
\end{center}
\caption{Renormalization scale as a function of $\sqrt{s}$ obtained from the constraint $\mbox{Abs}[ (\sigma_{\NLO} - \sigma_{\LO})/\sigma_{\LO}] = z$ with $z$ taken in the range $0.1 \div 0.5$. Left plot is for PP final state with both $\gamma^*$ and $Z^*$ contributions taken into account (results for PV and VV are almost the same as for PP, so they were omitted here); solid line corresponds to $\mu = \sqrt{s}$. Right plot is for PV combination of final mesons with account for only $\gamma^*$ contribution.
\label{fig_scale_choice}}
\end{figure}

To investigate the dependence of full NLO cross section on the choice of renormalization scale and to deduce a range of admissible values  we employed the following strategy. We defined the measure of importance of $O(\alpha_S^3)$ corrections as:
\begin{equation}
\label{eq_def_R}
\mathcal R = \mbox{Abs}\left[ \frac{\sigma_{\NLO} - \sigma_{\LO}}{\sigma_{\LO}}\right].
\end{equation}
In order to have perturbation series ``convergent''\footnote{Here we require that NLO contribution is not higher than LO result, otherwise of course perturbation series is asymptotic.} $\mathcal R$ should be small in all energy range. So, we first adopted the constraint $\mathcal R = z$ for a set of fixed $z$ values in the range $\left[0.1 \div 0.5\right]$ and solved the corresponding equation with respect to $\mu$. Fig.~\ref{fig_scale_choice} (left) shows obtained renormalization scales $\mu$ as a function of $\sqrt{s}$ in the case of production of a pair of pseudoscalar mesons (results for other states are the same and we omitted them in the figure). It is seen, that if one takes $\mu$ proportional to $\sqrt{s}$, $\mathcal R$ will be constant in all energy range. In particular, taking $\mu = \sqrt{s}$ we have $\mathcal R = 0.3$, while varying $\mu$ in the range $ \sqrt{s}/2 \leq \mu \leq 2 \sqrt{s}$, $\mathcal R$ varies in the range $0.2 \leq \mathcal R \leq 0.5$. This result is quite predictable, since it is common to take renormalization scale equal to some characteristic energy of the process and in the case of the considered processes $\sqrt{s}$ appears in all propagators. It is interesting to note, that in the case of production of pseudoscalar plus vector (PV) mesons without account for virtual $Z^*$ boson the situation is completely different. Fig.~\ref{fig_scale_choice} (right) shows, that in this case the renormalization scale, which ensures smallness of $\mathcal R$ value at all energies, is inversely proportional to $\sqrt{s}$. This is probably related to the fact that in this case there is an additional $M/\sqrt{s}$ suppression in the amplitude as discussed previously.

\begin{figure}
\begin{center}
\includegraphics[width=0.45\textwidth]{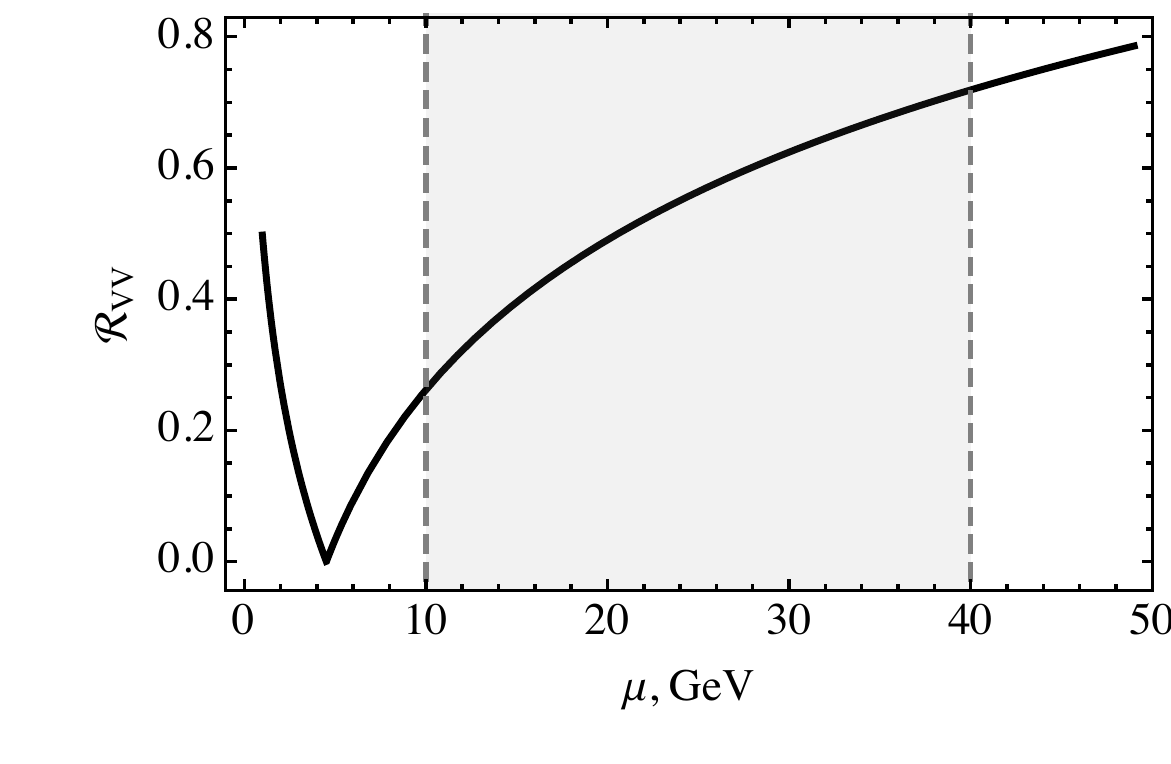} 
\hspace{0.02\textwidth} 
\includegraphics[width=0.45\textwidth]{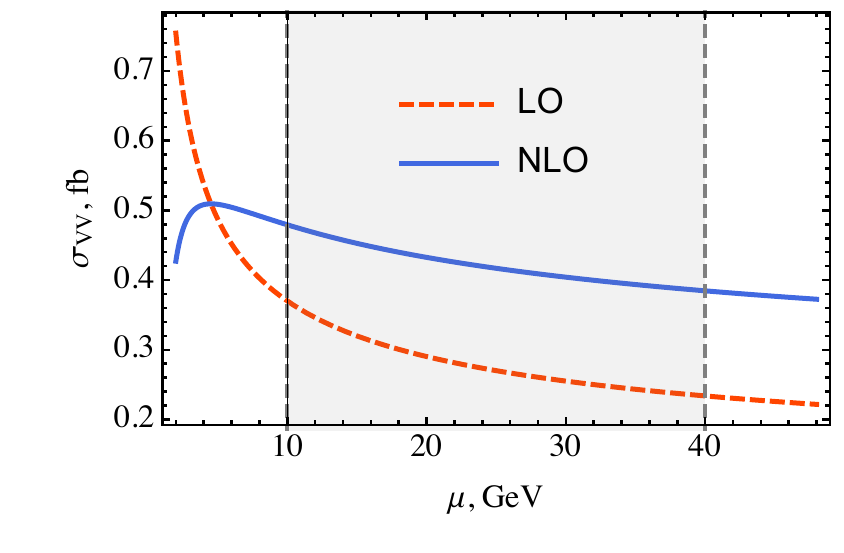}  
\end{center}
\caption{Scale dependence of the NLO to LO ratio $\mathcal R$ defined in \eqref{eq_def_R} (left) and the total cross section (right). Both plots are evaluated for VV final state at $\sqrt{s} = 20\,\GeV$ (results for PP and PV are almost the same and thus they were omitted). Filled areas correspond to interval $\sqrt{s}/2 \leq \mu \leq 2 \sqrt{s}$. Two curves on the right plot correspond to leading order (dashed curve) and full NLO cross section (solid curve).
\label{fig_scale_choice_fixed_ecm}}
\end{figure}

\begin{figure}
\begin{center}
\includegraphics[width=0.45\textwidth]{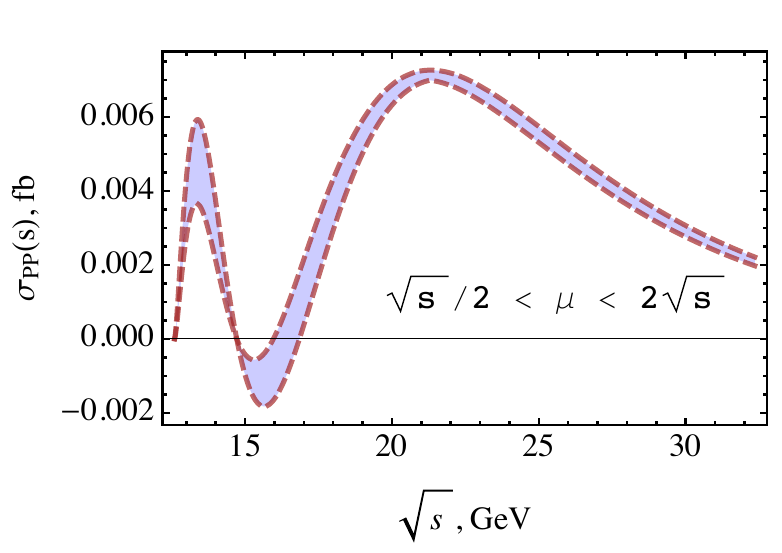} 
\hspace{0.02\textwidth} 
\includegraphics[width=0.45\textwidth]{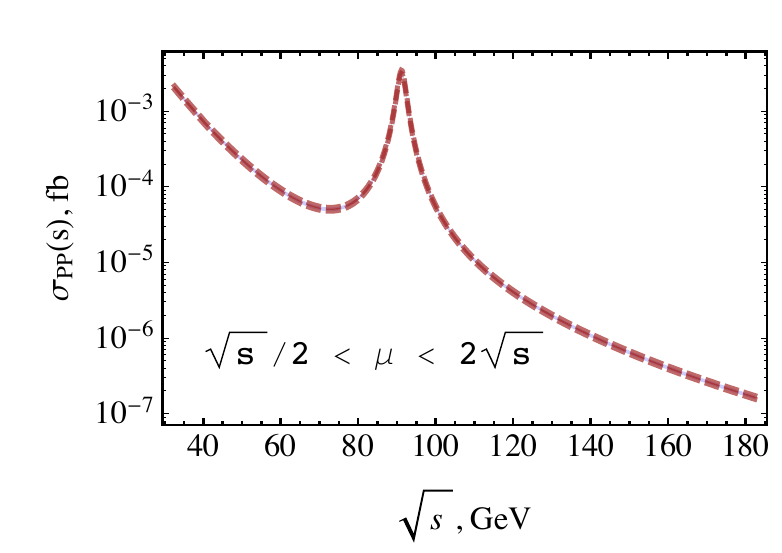}  \\
\includegraphics[width=0.45\textwidth]{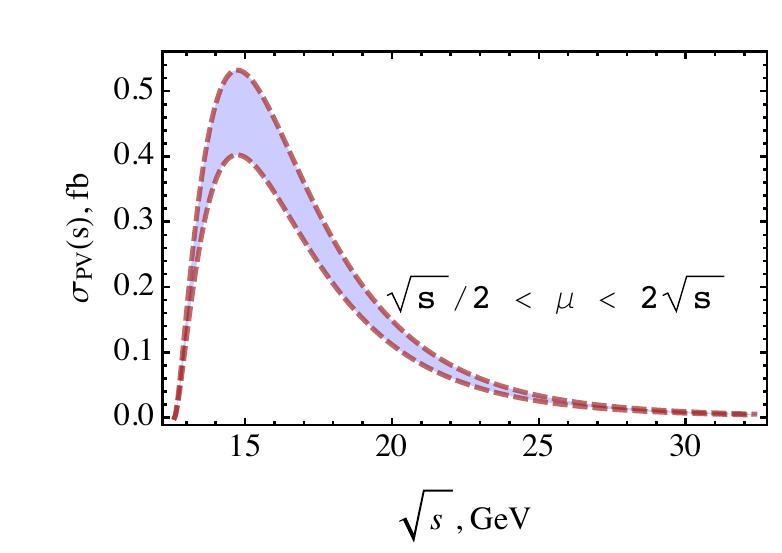} 
\hspace{0.02\textwidth} 
\includegraphics[width=0.45\textwidth]{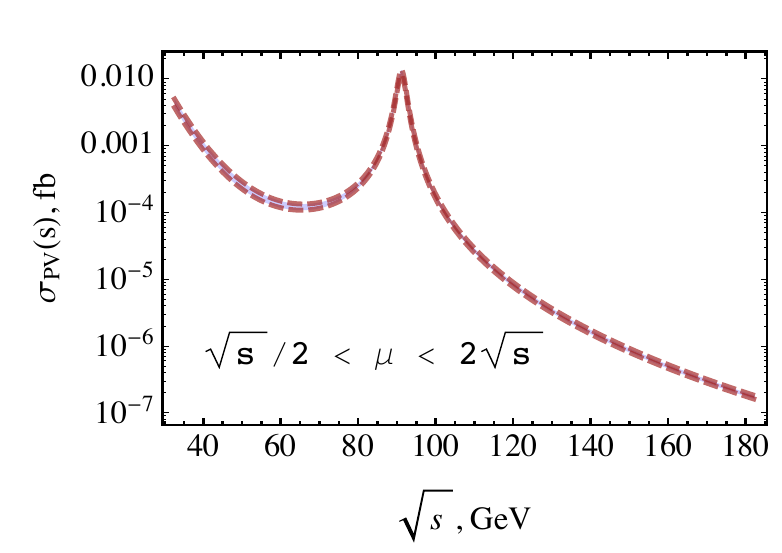}  \\
\includegraphics[width=0.45\textwidth]{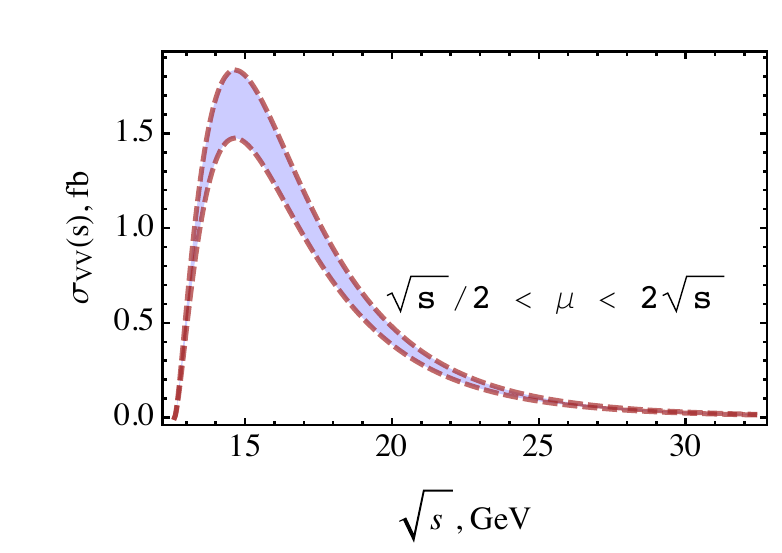} 
\hspace{0.02\textwidth} 
\includegraphics[width=0.45\textwidth]{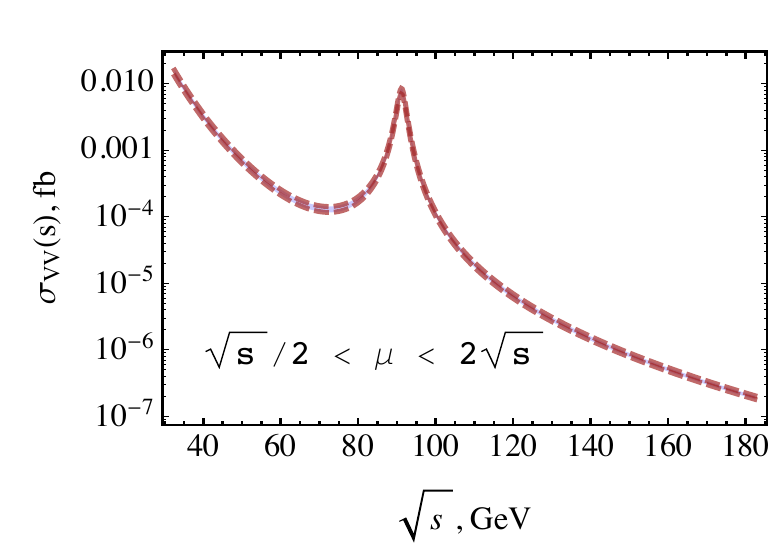} 
\hspace{0.02\textwidth} 
\end{center}
\caption{Dependence on renormalization scale $\mu$ evaluated by varying $\mu$ within the interval $\sqrt{s}/2 \leq \mu \leq 2 \sqrt{s}$. Upper plots correspond to the production of a pair of pseudoscalar (PP) mesons, middle for pseudoscalar and vector mesons (PV) and lower for a pair of vector mesons (VV).
\label{fig_nlo_scale}}
\end{figure}

Varying $\mu$ value in the range $\sqrt{s}/2 \leq \mu \leq 2\sqrt{s}$ the NLO contribution for the production of different states varies as
\begin{eqnarray*}
&& 0.2 \leq \mathcal R_{\mPP} \leq 0.5, \quad
0.3 \leq \mathcal R_{\mPV} \leq 0.9,\quad
0.3 \leq \mathcal R_{\mVV} \leq 0.7.
\end{eqnarray*}
Fig.~\ref{fig_scale_choice_fixed_ecm} (left) shows how $\mathcal R$ value depends on $\mu$ at a fixed energy scale  $\sqrt{s} = 20\,\GeV$ 
in the case of the production of a pair of vectors (VV) mesons (the results for another states and energies are similar). Fig.~\ref{fig_scale_choice_fixed_ecm} (right) shows this dependence for LO and NLO cross sections. It is seen, that the variation of  $\mu$ leads to 35\% uncertainty in the LO cross section and only to 10\% uncertainty for the NLO cross section. So, as expected an account for NLO corrections stabilizes the dependence of cross section on $\mu$ and provides higher precision of theoretical predictions. Finally, Tab.~\ref{tab_scales} presents cross sections obtained at different energies and Fig.~\ref{fig_nlo_scale} shows results for NLO cross sections with the uncertainties induced by the variation of $\mu$.

\begin{table}
	\begin{center}
		\begin{tabular}{|c||c|c|c|c|c|}
			\hline
			&  & $\mu = \sqrt{s}/2$ & $\mu = \sqrt{s}$ & $\mu = 2\sqrt{s}$ & $\mu = 2M_{B_c}$\\
			\hline\hline
			\multirow{3}{*}{$\sqrt{s} = 17\, \GeV$} 
			& PP & $3.8\times 10^{-4}$ fb & $1.6\times 10^{-3}$ fb & $2.2\times 10^{-3}$ fb & $1.2\times 10^{-3}$ fb\\
			
			& PV & $3.6\times 10^{-1}$ fb & $3.1\times 10^{-1}$ fb & $2.7\times 10^{-1}$ fb & $3.3\times 10^{-1}$ fb\\
			
			& VV & $1.2$ fb & $1.1$ fb & $9.6\times 10^{-1}$ fb & $1.1$ fb\\
			\hline
			\multirow{3}{*}{$\sqrt{s} = m_Z$} 
			& PP & $3.5\times 10^{-3}$ fb & $3.3\times 10^{-3}$ fb & $3.1\times 10^{-3}$ fb & $3.1\times 10^{-3}$ fb\\
			
			& PV & $1.3\times 10^{-2}$ fb & $1.2\times 10^{-2}$ fb & $1.1\times 10^{-2}$ fb & $1.4\times 10^{-2}$ fb\\
			
			& VV & $8.4\times 10^{-3}$ fb & $7.7\times 10^{-3}$ fb & $6.9\times 10^{-3}$ fb & $9.5\times 10^{-3}$ fb\\
			\hline
			\multirow{3}{*}{$\sqrt{s} = 2 \times m_Z$} 
			& PP & $1.7\times 10^{-7}$ fb & $1.6\times 10^{-7}$ fb & $1.5\times 10^{-7}$ fb & $1.4\times 10^{-7}$ fb\\
			
			& PV & $1.8\times 10^{-7}$ fb & $1.7\times 10^{-7}$ fb & $1.6\times 10^{-7}$ fb & $1.6\times 10^{-7}$ fb\\
			
			& VV & $2.\times 10^{-7}$ fb & $1.9\times 10^{-7}$ fb & $1.8\times 10^{-7}$ fb & $1.7\times 10^{-7}$ fb\\
			\hline
		\end{tabular}
	\end{center}
	\caption{ cross sections evaluated at different renormalization scales.
		\label{tab_scales}}
\end{table}

\begin{figure}
\begin{center}
\includegraphics[width=0.45\textwidth]{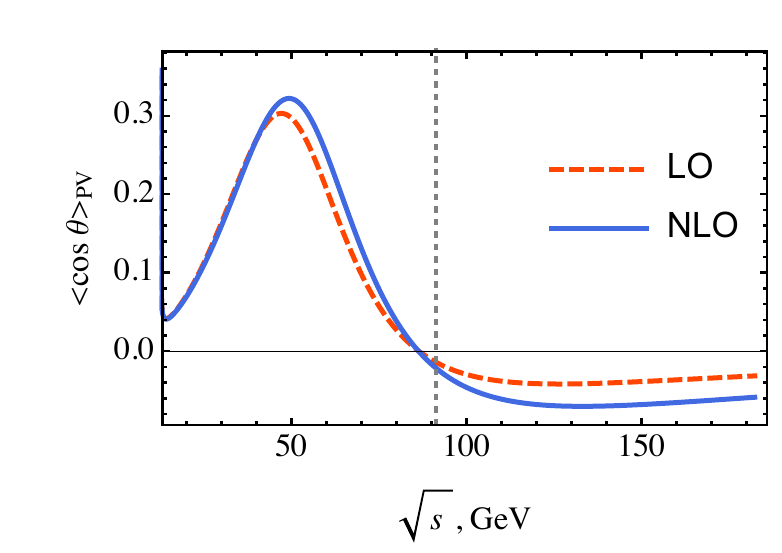} 
\hspace{0.02\textwidth} 
\includegraphics[width=0.45\textwidth]{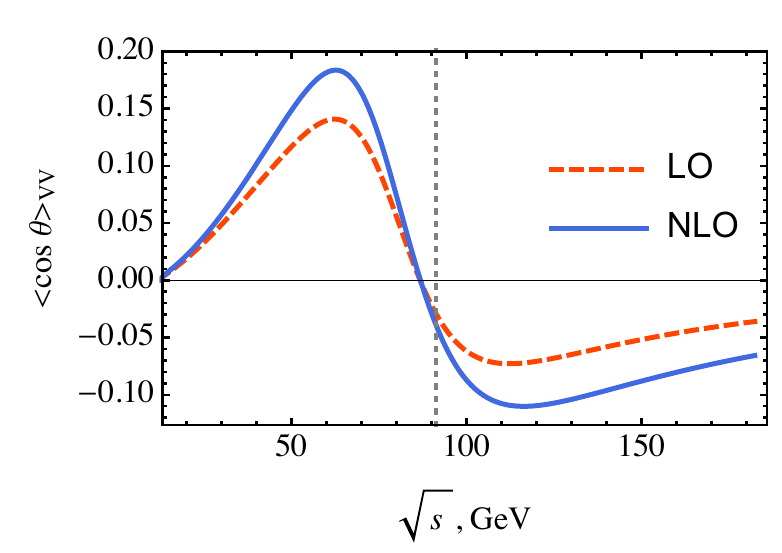}
\end{center}
\caption{Azimuthal asymmetry for PV (left) and VV (right) mesons production. Dashed curves correspond to LO contributions, solid for the full NLO result. Vertical dotted line corresponds to $\sqrt{s} = m_Z$.
\label{fig_asymmetry_ecm}}
\end{figure}

\begin{figure}
\begin{center}
\includegraphics[width=0.45\textwidth]{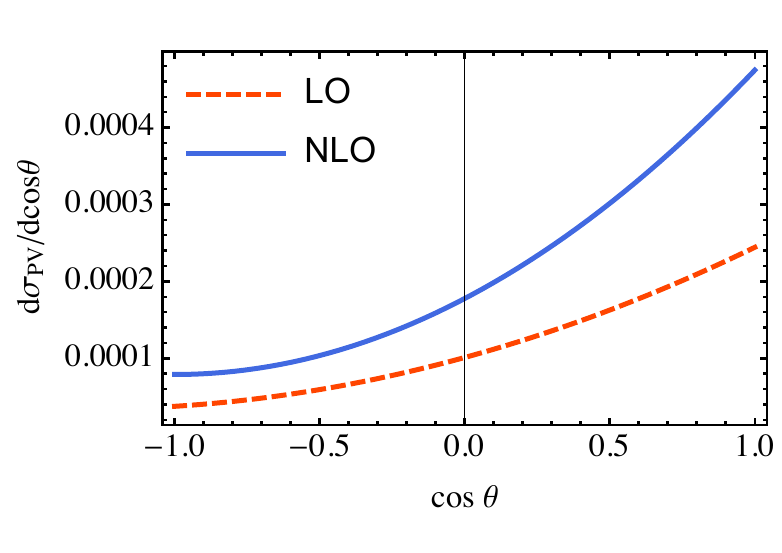} 
\hspace{0.02\textwidth} 
\includegraphics[width=0.45\textwidth]{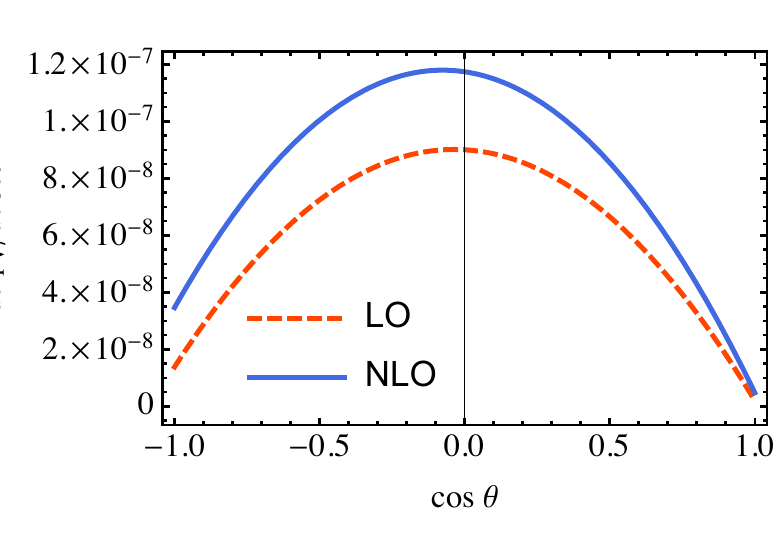}  \\
\includegraphics[width=0.45\textwidth]{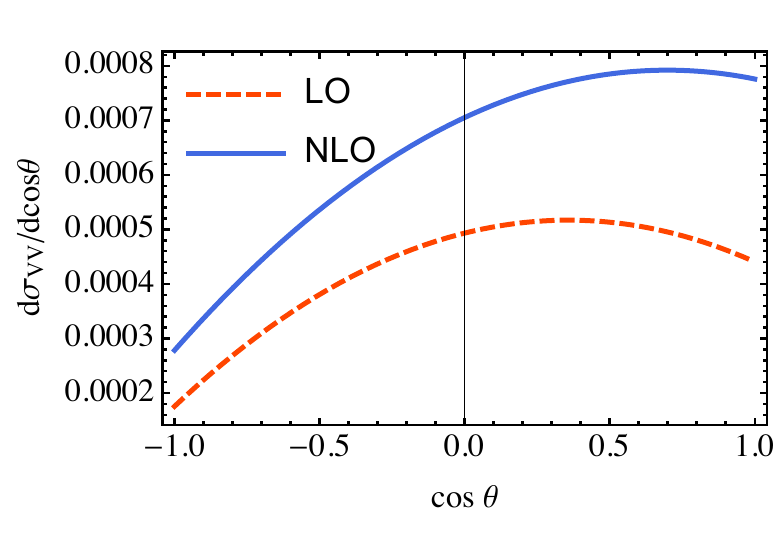} 
\hspace{0.02\textwidth} 
\includegraphics[width=0.45\textwidth]{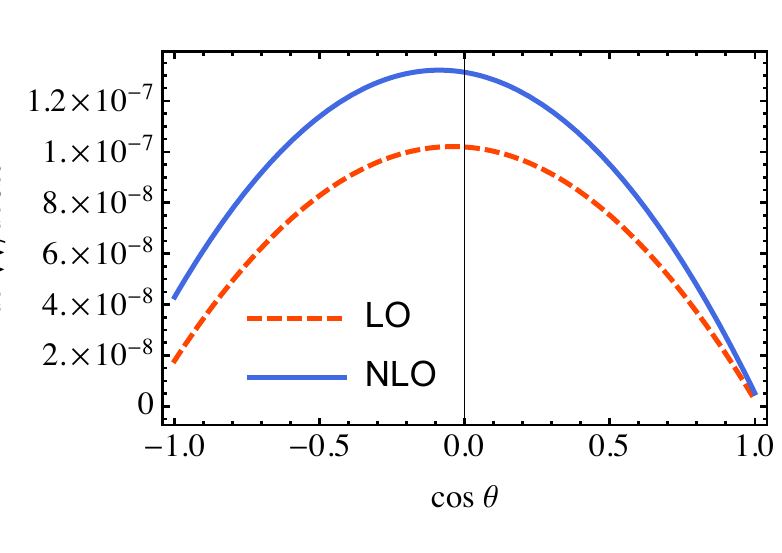}
\end{center}
\caption{Differential cross section $d\sigma/d\cos\theta$ at $\sqrt{s} = m_Z/2$ (left) and $\sqrt{s} = 2 m_Z$ (right). Upper plots correspond to the production of a pair of pseudoscalar and vector (PV) mesons, while lower for a pair of vectors (VV) mesons.
\label{fig_asymmetry_cos}}
\end{figure}

Weak interactions are known to be $P$-violating and thus we should observe this phenomenon in the considered processes. To study $P$-asymmetry we choose the following quantity:
$$
\mathcal A \equiv \left \langle \cos(\theta) \right \rangle = \frac{1}{\sigma}\, \int d \cos(\theta) \left( \cos(\theta) \, \frac{d\sigma}{d\cos(\theta)} \right),
$$
which is of the order of the difference $\left(\sigma(\theta = \pi) - \sigma(\theta = -\pi)\right)$. In the case of PP production this asymmetry is  exactly zero in all energy range. This is easily explained by the fact that axial part of $Z$-current which is $1^+$ can not produce two $0^-$ states and thus axial current gives no contribution in this case and there is no V-A interference and parity violation. In the case of PV and VV mesons production the dependence of $\mathcal A$ on the center of mass energy is shown in Fig.~\ref{fig_asymmetry_ecm}. It is seen, that the maximal asymmetry should be observed at $\sqrt{s} \approx m_Z/2$ and it is almost zero at $Z$-boson peak. Specifically, the asymmetry becomes zero at the following points in the case of  PV and VV mesons production:
\begin{eqnarray*}
&& s_{\mPV} = m_Z^2\frac{2 (1-3 r ((r-1) r+1)) \sin ^2\left(2 \theta _W\right)}{6 r^3 \cos \left(2 \theta _W\right)+(1-3 r ((r-1) r+1)) \cos \left(4 \theta _W\right)-3 (r (3 r-2)+2) r+2} \approx 0.9 \, m_Z^2,\\
&& s_{\mVV} = m_Z^2 \frac{2 (r (r (r+3)-3)+1) \sin ^2\left(2 \theta _W\right)}{-6 r^3 \cos \left(2 \theta _W\right)+(r (r (r+3)-3)+1) \cos \left(4 \theta _W\right)+(r (5 r+6)-6) r+2} \approx 0.9 \, m_Z^2.
\end{eqnarray*}
Fig.~\ref{fig_asymmetry_cos} shows $d\sigma/d\cos(\theta)$ for the case of PV and VV mesons production at $\sqrt{s} = m_Z/2$ and $\sqrt{s} = 2 m_Z$.

\section{Conclusions}
In the present paper we have considered paired production of different $S$-wave $B_c$ states in  $e^+e^-$-annihilation via virtual $\gamma^*$ and $Z^*$ bosons taking into account full NLO QCD corrections. We have found that one-loop corrections are quite sizable and may raise the LO cross section up to 2 times at low energies, while at high energies $\sigma_{\NLO}/\sigma_{\LO} \approx 1.2$. The dependence on the renormalization scale $\mu$ stabilizes with the account of NLO corrections, so that errors induced by variation of the scale are reduced from 30\% to only 10\% at NLO compared to LO.

An interesting feature of the production cross section for PP state is that it becomes zero at some energy point \eqref{eq_zero_PP_point} near threshold ($\sqrt{s_{\mPP}} \approx 14.7\,\GeV$). One-loop contribution at this point makes cross section negative and thus both two-loop contribution and account for relativistic corrections are required in order to correctly describe PP mesons production near threshold. At the same time, cross sections for PV and especially VV mesons production have maxima at this energy. Thus, the production of $0^-$ states is strongly suppressed at this point, while the production of $1^-$ states has significant cross section. The latter fact provides us with an excellent setup for a possible search and studies of $B_c^*$. 

Account of intermediate $Z^*$-boson is important for the description of the production processes near $Z$-boson peak. Additionally, in the case of PV mesons production an account of $Z$-boson exchange changes asymptotic behavior of the cross section at high energies from $1/s^4$ to $1/s^3$ --- additional $1/s$ suppression in the case of vector current only can be explained by approximate chirality conservation. Another interesting feature related with this is that account for $Z$-boson completely changes PV cross section dependence on the renormalization scale $\mu$. 

Finally, parity asymmetry induced by weak current appears to be most significant near $\sqrt{s} = m_Z/2$ point, while it is almost zero at $\sqrt{s} = m_Z$. It is interesting to note, that in the case of PP mesons production there is no parity asymmetry since a pair of $0^-$ states can not be produced by axial current thus there is no V-A interference in the cross section.

\section*{Acknowledgements}
 We are grateful to T. Aushev, A. Bednyakov,  V. Shtabovenko and H.Patel for help and fruitful discussion. This work was supported by  Russian Foundation of Basic Research (grants \#15-02-03244 and \#14-02-00096) and contract \# 02.A03.21.0003 from 27.08.2013 with Russian Ministry of Science and Education. The work of S.V. Poslavsky was supported by the Russian Foundation of Basic Research (grant \#16-32-60017).  A.V. Berezhnoy acknowledges the support from MinES of RF (grant 14.610.21.0002, identification number RFMEFI61014X0002).

\appendix
\section{Leading order cross sections}
\label{appendixA}

In the following we use the notation: $M = m_c + m_b$, $r = m_c/M$ and $s = (k_1 + k_2)^2/M^2$, where $k_1$ and $k_2$ are incoming momenta of electron-positron pair. The cross section for PP mesons production is given by:
\begin{eqnarray}
\frac{d\sigma_{\mPP}}{d\cos(\theta)} = \frac{\pi^3 \alpha^2 \alpha_S^2 f_{B_c}^4 (s-4)^{3/2}}{729 M^2 (r-1)^6 r^6 s^{9/2}} \left( \sigma_{\mPP}^{\gamma\gamma} + \sigma_{\mPP}^{\gamma Z} + \sigma_{\mPP}^{ZZ} \right),
\end{eqnarray}
where $\sigma_{\mPP}^{\gamma\gamma}$, $\sigma_{\mPP}^{\gamma Z}$ and $\sigma_{\mPP}^{ZZ}$ are defined as
\begin{eqnarray}
&& \sigma_{\mPP}^{\gamma\gamma}  = \frac{16 \left(1-\cos^2(\theta)\right) a_{\mPP}^2}{M^4 s^2},\\
&& \sigma_{\mPP}^{\gamma Z}  =-\frac{2 \left(1-\cos^2(\theta)\right) a_{\mPP} b_{\mPP} \left(\csc \left(\theta_W\right)-2\right) \left(\csc \left(\theta_W\right)+2\right) \sec^2\left(\theta_W\right) \left(M^2 s-m_Z^2\right)}{M^2 s \left(\left(m_Z^2-M^2 s\right){}^2+\Gamma^2 m_Z^2\right)},\\
&& \sigma_{\mPP}^{ZZ}  = \frac{\left(1-\cos^2(\theta)\right) b_{\mPP}^2 \left(-2 \cos \left(2 \theta_W\right)+\cos \left(4 \theta_W\right)+2\right) \csc^4\left(\theta_W\right) \sec^4\left(\theta_W\right)}{8 \left(\left(m_Z^2-M^2 s\right){}^2+\Gamma^2 m_Z^2\right)}.
\end{eqnarray}
Here
\begin{eqnarray}
&& a_{\mPP} = r \left(-3 r^3 (s+2)+r^2 (5 s+8)-3 r (s+4)+s+8\right)-2,\\
&& b_{\mPP} =  2 (r (r (r (3 r (s+2)-5 s-8)+3 (s+4))-s-8)+2) \cos \left(2 \theta_W\right) \\ 
&&\nonumber \phantom{b_{\mPP} ===} +\,(1-2 r) ((r-1) r (s+4)+2).
\end{eqnarray}

The cross section for PV mesons production is
\begin{eqnarray}
\frac{d\sigma_{\mPV}}{d\cos(\theta)} = \frac{\pi^3 \alpha^2 \alpha_S^2 f_{B_c}^4 (s-4)}{243 M^2 (r-1)^6 r^6 s^{4}} \left( \sigma_{\mPV}^{\gamma\gamma} + \sigma_{\mPV}^{\gamma Z} + \sigma_{\mPV}^{ZZ} \right),
\end{eqnarray}
where $\sigma_{\mPV}^{\gamma\gamma}$, $\sigma_{\mPV}^{\gamma Z}$ and $\sigma_{\mPV}^{ZZ}$ are defined as
\begin{eqnarray}
&& \sigma_{\mPV}^{\gamma\gamma}  = \frac{16 \sqrt{s-4} \left(1+\cos^2(\theta)\right) a_{\mPV}^2}{3 M^4 s^{3/2}},\\
&& \sigma_{\mPV}^{\gamma Z}  = \frac{16 a_{\mPV} \csc^2\left(2 \theta_W\right) \left(M^2 s-m_Z^2\right) }{M^2 s \left(\left(m_Z^2-M^2 s\right){}^2+\Gamma^2 m_Z^2\right)} \left(c_{\mPV} \cos (\theta )  -\frac{1}{6} \sqrt{s^2-4 s} \left(\cos^2(\theta )+1\right) \times \right.\\
&&\nonumber \hspace{4cm}\left. 
\times \left(2 \cos \left(2 \theta_W\right)-1\right) \left(-2 a_{\mPV} \cos \left(2 \theta_W\right)-3 r^2+3 r-1\right)\right),\\
&& \sigma_{\mPV}^{ZZ}  =
\frac{\csc^4\left(\theta_W\right) \sec^4\left(\theta_W\right) }{2 \left(\left(m_Z^2-M^2 s\right){}^2+\Gamma^2 m_Z^2\right)} \\
&&\nonumber\hspace{1cm}
\left(-\frac{1}{12} \sqrt{1-\frac{4}{s}} \cos^2(\theta ) \left(-2 \cos \left(2 \theta_W\right)+\cos \left(4 \theta_W\right)+2\right) \times \right. \\
&&\nonumber\hspace{3cm} \times \,
 \left(-2 s a_{\mPV}^2 \cos \left(4 \theta_W\right)+4 s b_{\mPV} \cos \left(2 \theta_W\right)+d_{\mPV}\right)\\
&&\nonumber\hspace{1.5cm}  
 +\frac{1}{12 \sqrt{s^2-4 s}}\left(-2 \cos \left(2 \theta_W\right)+\cos \left(4 \theta_W\right)+2\right) \times \\ 
 &&\nonumber\hspace{3cm}   \times\,
 \left(2 (s-4) s a_{\mPV}^2 \cos \left(4 \theta_W\right)-4 (s-4) s b_{\mPV} \cos \left(2 \theta_W\right)+e_{\mPV}\right)\\
&&\nonumber\hspace{1.5cm} \left.
 \phantom{\sqrt{\frac{1}{2}}}-c_{\mPV} \cos (\theta ) \left(2 \cos \left(2 \theta_W\right)-1\right) \left(-2 a_{\mPV} \cos \left(2 \theta_W\right)-3 r^2+3 r-1\right)\right).
\end{eqnarray}
Here
\begin{eqnarray}
&& a_{\mPV} = 1-3 r ((r-1) r+1),\\
&& b_{\mPV} =  (3 (r-1) r+1) (3 r ((r-1) r+1)-1),\\
&& c_{\mPV} =  2 (r-1) r ((r-1) r (s+4)-s+8)-s+4,\\
&& d_{\mPV} = 9 (r-1)^2 r^2 (2 (r-1) r+1)^2 s^2\\
&& \nonumber \phantom{d_{\mPV} =}+3 (r (r (r (3 r (2 r (r (6 (r-4) r+47)-58)+81)-74)-15)+18)-4) s\\
&& \nonumber \phantom{d_{\mPV} =}+36 (2 (r-1) r ((r-1) r+2)+1)^2,\\
&& e_{\mPV} = 3 (r (r (r (3 r (2 r (r (2 (r-4) r+9)+2)-13)+2)+27)-18)+4) s^2\\
&& \nonumber \phantom{d_{\mPV} =} +  9 (r-1)^2 r^2 (2 (r-1) r+1)^2 s^3 +144 (2 (r-1) r ((r-1) r+2)+1)^2 \\
&& \nonumber \phantom{d_{\mPV} =} +12 (r (r (r (3 r (2 r (r (4 (r-4) r+27)-26)+9)+94)-99)+42)-7) s.
\end{eqnarray}

The cross section for VV mesons production is
\begin{eqnarray}
\frac{d\sigma_{\mVV}}{d\cos(\theta)} = \frac{\pi^3 \alpha^2 \alpha_S^2 f_{B_c}^4 (s-4)^{3/2}}{243 M^2 (r-1)^6 r^6 s^{9/2}} \left( \sigma_{\mVV}^{\gamma\gamma} + \sigma_{\mVV}^{\gamma Z} + \sigma_{\mVV}^{ZZ} \right),
\end{eqnarray}
where $\sigma_{\mVV}^{\gamma\gamma}$, $\sigma_{\mVV}^{\gamma Z}$ and $\sigma_{\mVV}^{ZZ}$ are defined as
\begin{eqnarray}
&& \sigma_{\mVV}^{\gamma\gamma}  = \frac{16 \left(b_{\mVV}-a_{\mVV} \cos^2(\theta )\right)}{3 M^4 s^2},\\
&& \sigma_{\mVV}^{\gamma Z}  =\frac{16 \csc^2\left(2 \theta_W\right) \left(M^2 s-m_Z^2\right) }{M^2 s \left(\left(m_Z^2-M^2 s\right){}^2+\Gamma^2 m_Z^2\right)} \\
&&\nonumber\hspace{1cm}  \left(-\frac{1}{6} \cos^2(\theta ) \left(2 \cos \left(2 \theta_W\right)-1\right) \left(2 a_{\mVV} \cos \left(2 \theta_W\right)+c_{\mVV}\right) \right.\\
&&\nonumber\hspace{1.5cm} 
+\, \frac{1}{6} \left(2 \cos \left(2 \theta_W\right)-1\right) \left(2 b_{\mVV} \cos \left(2 \theta_W\right)+d_{\mVV}\right)\\
&&\nonumber\hspace{1.5cm} \left. 
+\, 2  (2 r-1) \left(2 r^2-2 r+1\right) \left(r^3+3 r^2-3 r+1\right) s \sqrt{1-\frac{4}{s}} \cos (\theta )\right),\\
&& \sigma_{\mVV}^{Z Z}  = \frac{\csc^4\left(\theta_W\right) \sec^4\left(\theta_W\right) }{\left(m_Z^2-M^2 s\right){}^2+\Gamma^2 m_Z^2}\\
&&\nonumber\hspace{1cm} 
 \left(-\frac{1}{24} \cos^2(\theta ) \left(-2 \cos \left(2 \theta_W\right)+\cos \left(4 \theta_W\right)+2\right) \times \right. \\
&&\nonumber\hspace{3cm}   \times\,
 \left(2 a_{\mVV} \cos \left(4 \theta_W\right)+4 c_{\mVV} \cos \left(2 \theta_W\right)+e_{\mVV}\right) \\
&&\nonumber\hspace{1cm}   \,
 +\,\frac{1}{24} \left(-2 \cos \left(2 \theta_W\right)+\cos \left(4 \theta_W\right)+2\right) \times\\
 &&\nonumber\hspace{3cm} \times\,
 \left(2 b_{\mVV} \cos \left(4 \theta_W\right)+4 d_{\mVV} \cos \left(2 \theta_W\right)+f_{\mVV}\right)\\
&&\nonumber\hspace{1cm}  \,
 +\,\left(4 r^3-6 r^2+4 r-1\right) \sqrt{s^2-4 s} \cos (\theta ) \left(2 \cos \left(2 \theta_W\right)-1\right) \times \\
 &&\nonumber\hspace{3cm} \left.
\phantom{\frac{1}{24}} \times \, \left(-2 r^3+3 r^2+2 \left(r^3+3 r^2-3 r+1\right) \cos \left(2 \theta_W\right)-3 r+1\right)\right).
\end{eqnarray}
Here
\begin{eqnarray}
&& a_{\mVV} = 12 (r (r (r (3 r-4)+6)-4)+1)^2 + (r-1)^2 r^2 (r (3 r-2)+1)^2 s^2\\
&&\nonumber \phantom{a_{\mVV} =} -2 (r (r (r (r (r (r (18 (r-3) r+95)-108)+93)-62)+29)-8)+1) s,\\
&& b_{\mVV} = (r-1)^2 r^2 (r (3 r-2)+1)^2 s^2+12 (r (r (r (3 r-4)+6)-4)+1)^2\\
&&\nonumber \phantom{a_{\mVV} =} +2 \left(r \left(-18 r^7+54 r^6-93 r^5+120 r^4-87 r^3+30 r^2+r-4\right)+1\right) s,\\
&& c_{\mVV} = (1-2 r) \left(r \left(r \left((r-1)^2 (r (3 r-2)+1) s^2+2 r (r ((22-9 r) r-31)+28) s \right.\right.\right. \\
&&\nonumber \phantom{a_{\mVV} =} \left.\left.\left.+12 r (r (2 r (3 r-7)+23)-24)-34 s+192\right)+12 (s-6)\right)-2 (s-6)\right),\\
&& d_{\mVV} = (1-2 r) \left(r \left(r \left((r-1)^2 (r (3 r-2)+1) s^2-6 ((r-1) r+1) (r (3 r-5)+1) s\right.\right.\right. \\
&&\nonumber \phantom{a_{\mVV} =} \left.\left.\left.+12 (r (r (2 r (3 r-7)+23)-24)+16)\right)-4 (s+18)\right)+2 (s+6)\right),\\
&& e_{\mVV} = 3 \left(r \left(r \left((r-1)^2 (2 r (r (r (3 r-4)+4)-2)+1) s^2\right.\right.\right. \\
&&\nonumber \phantom{a_{\mVV} =} -2 (r (r (2 r (r (6 (r-3) r+59)-118)+293)-230)+113) s \\
&&\nonumber \phantom{a_{\mVV} =} \left.\left.\left.+24 (r (r (r (r (r (3 r-8)+36)-80)+104)-84)+42)\right)+32 (2 s-9)\right)-8 s+36\right),\\
&& f_{\mVV} = 3 \left(r \left(r \left(24 r \left((r-2) r \left((3 r-2) r^2+16\right)+28\right) \right.\right.\right. \\
&&\nonumber \phantom{a_{\mVV} =} +(r-1)^2 (2 r (r (r (3 r-4)+4)-2)+1) s^2\\
&&\nonumber \phantom{a_{\mVV} =} \left.\left.\left.-2 r (r (2 r (3 r (2 (r-3) r+3)+26)-133)+142) s+170 s-336\right)-56 s+96\right)+8 s-12\right).
\end{eqnarray}

\section{$A_0$ and $B_0$ master integrals}
\label{appendixB}

This Appendix contains analytical expressions for $A_0$ and $B_0$ master integrals up to $\mathcal{O}(\epsilon^2)$ order in $\epsilon$-expansion. They are given by \cite{Davydychev:2000na,Berends:1996gs} ( $D=4-2\epsilon , s = q^2$ ):
\begin{eqnarray}
A_0 (m) &\equiv& \int\frac{d^D k}{(2\pi)^D}\frac{1}{k^2 - m^2} 
= -\frac{i\pi^{2-\epsilon}}{(2\pi)^{4-2\epsilon}}\Gamma (-1+\epsilon) (m^2)^{1-\epsilon} ,  \\
B_0 (s,0,0) &\equiv& \int\frac{d^D k}{(2\pi)^D}\frac{1}{k^2 (k-q)^2} = \frac{i\pi^{2-\epsilon}}{(2\pi)^{4-2\epsilon}} (-s)^{-\epsilon}
\frac{\Gamma^2 (1-\epsilon)\Gamma (\epsilon)}{\Gamma (2-2\epsilon)} , \\ 
B_0 (s,m,0) &\equiv& \int\frac{d^D k}{(2\pi)^D}\frac{1}{(k^2-m^2) (k-q)^2} = \frac{i\pi^{2-\epsilon}}{(2\pi)^{4-2\epsilon}} m^{-2\epsilon} \frac{\Gamma (1+\epsilon)}{\epsilon (1-\epsilon)} ~_2 F_1 \left(\begin{matrix}
1, \epsilon \\ 2-\epsilon
\end{matrix} ~\Big|~\frac{s}{m^2} \right) , \nonumber \\ \\
B_0 (s,m_1,m_2) &\equiv& \int\frac{d^D k}{(2\pi)^D}\frac{1}{(k^2-m_1^2) ((k-q)^2 - m_2^2)} = \frac{i\pi^{2-\epsilon}}{(2\pi)^{4-2\epsilon}} (-s)^{-\epsilon} \Bigg\{ \frac{\Gamma^2 (1-\epsilon)\Gamma (\epsilon)}{\Gamma (2-2\epsilon)}\lambda^{1-2\epsilon} \nonumber \\
&& - \Gamma (-1+\epsilon) (1+x-y-\lambda)\frac{(-x)^{-\epsilon}}{2}
~_2 F_1 \left(\begin{matrix}
1, \epsilon \\ 2-\epsilon
\end{matrix} ~\Bigg|~\frac{(1+x-y-\lambda)^2}{4 x} \right) \nonumber \\
&& - \Gamma (-1+\epsilon) (1-x+y-\lambda)\frac{(-y)^{-\epsilon}}{2}
~_2 F_1 \left(\begin{matrix}
1, \epsilon \\ 2-\epsilon
\end{matrix} ~\Bigg|~\frac{(1-x+y-\lambda)^2}{4 y} \right) \Bigg\}
\end{eqnarray}
where 
\begin{eqnarray}
x = \frac{m_1^2}{s},\quad y = \frac{m_2^2}{s},\quad \lambda = \sqrt{(1-x-y)^2 - 4 x y }
\end{eqnarray}
and up to $\mathcal{O}(\epsilon^2)$ terms we have
\begin{eqnarray}
~_2 F_1 \left(\begin{matrix}
1, \epsilon \\ 2-\epsilon
\end{matrix} ~\Bigg|~z \right) = \frac{1-\epsilon}{2 (1-2\epsilon) z}
\left(1+ z - (1-z)^{1-2\epsilon} 
- 2 (1-z)^{1-2\epsilon}\epsilon^2 ~\text{Li}_2 (z) + \mathcal{O}(\epsilon^2)
\right) .
\end{eqnarray}

\section*{Bibliography}
\bibliographystyle{elsarticle-num}
\bibliography{litr}

\end{document}